# Cooling and heating with electron spins: Observation of the spin Peltier effect

J. Flipse*, F. L. Bakker*, A. Slachter, F. K. Dejene & B. J. van Wees

*Zernike Institute for Advanced Materials, Physics of Nanodevices, University of Groningen, 9747 AG Groningen, The Netherlands*

*These authors contributed equally to this work

**The Peltier coefficient describes the amount of heat that is carried by an electrical current when it passes through a material[1]. Connecting two materials with different Peltier coefficients causes a net heat flow towards or away from the interface, resulting in cooling or heating at the interface – the Peltier effect. Spintronics[2] describes the transport of charge and angular momentum by making use of separate spin-up and spin-down channels. Recently, the merger of thermoelectricity with spintronics has given rise to a novel and rich research field named spin caloritronics[3]. Here, we report the first direct experimental observation of refrigeration/heating driven by a spin current, a new spin thermoelectric effect which we call the spin Peltier effect. The heat flow is generated by the spin dependency of the Peltier coefficient inside the ferromagnetic material. We explored the effect in a specifically designed spin valve pillar structure by measuring the temperature using an electrically isolated thermocouple. The difference in heat flow between the two magnetic configurations leads to a change in temperature. With the help of 3-D finite element modeling, we extracted permalloy spin Peltier coefficients in the range of –0.9 to -1.3 mV. These results enable magnetic control of heat flow and provide new functionality for future spintronic devices.**

The coupling of heat transport with spintronics has generated novel ideas such as innovative spin sources[4-8], thermal spin-transfer torque[9,10], magnetic heat valves[11] and magnetically switchable cooling[12,13]. Driven by the downscaling of nanoelectronic components, the development and understanding of new and local refrigeration concepts is essential[14]. The spin Peltier effect offers this possibility, but direct experimental evidence of cooling by spin current has not yet been demonstrated. Pioneering experiments of Gravier *et al.* have reported field dependent magnetothermogalvanic voltage measurements in multiple Co/Cu multilayer nanowires, thereby indicating the existence of spin dependent Peltier coefficients[15]. Here we demonstrate the spin Peltier effect directly by electrically separating a temperature sensor from the spin Peltier source.

The spin Peltier effect is based on the assumption that spin-up and spin-down channels can transport heat independently. Figure 1a gives a schematic presentation of this concept when a pure spin current ($J_\uparrow = -J_\downarrow$) passes through a nonmagnetic metal / ferromagnetic metal (N/F) interface. The associated Peltier heat current $Q_\Pi$ is the sum of that of the two spin channels. Since both travel in opposite directions (the charge flow is zero), a net heat flow will only arise if the amount of heat carried by the separate spin species is different. The spin-dependent Peltier coefficients, defined as $\Pi_{\uparrow,\downarrow} = Q_{\Pi\uparrow,\downarrow} / J_{\uparrow,\downarrow}$, represent the amount of heat carried by the individual spin channels. In N, both coefficients are equal ($\Pi_\uparrow = \Pi_\downarrow$). However, in the ferromagnet the spin Peltier coefficient, defined as $\Pi_S = \Pi_\uparrow - \Pi_\downarrow$, is expected to be nonzero[12]. Owing to spin flip processes, the spin current attenuates in the ferromagnet and within a few spin relaxation lengths ($\lambda_F$) from the interface, the Peltier heat current vanishes. Consequently, heat is effectively transferred from the interface into the ferromagnetic region over a finite length, thereby producing a temperature gradient (depicted in Fig. 1b) and a temperature drop $\Delta T$.

In order to demonstrate the effect experimentally, a pure spin current is not required, but it can be accompanied by a charge current $J_C = J_\uparrow + J_\downarrow$. To calculate the temperature gradient, the total heat current $Q = Q_\Pi - \kappa \nabla T$ is evaluated where $\kappa$ is the thermal conductivity of the electron and phonon systems and $Q_\Pi = \Pi_\uparrow J_\uparrow + \Pi_\downarrow J_\downarrow$ is the Peltier heat current. If we assume that no heat can enter or leave the stack ($Q = 0$) and disregard Joule heating, the temperature gradient can be expressed as the sum of the charge and spin part of the Peltier effect (see Supplementary A). For the spin part, the induced temperature difference between the F/N interface and the bulk of the ferromagnet is given by:

$$\Delta T = \frac{\sigma}{4\kappa}\left(1 - P_\sigma^2\right)\Pi_S \mu_S^0 \qquad (1)$$

where $\sigma = \sigma_\uparrow + \sigma_\downarrow$ is the conductivity, $P_\sigma = \dfrac{\sigma_\uparrow - \sigma_\downarrow}{\sigma_\uparrow + \sigma_\downarrow}$ the conductivity polarization and $\mu_S^0 = \mu_\uparrow^0 - \mu_\downarrow^0$ the spin accumulation at the interface. Therefore, we find that the induced temperature drop depends directly on the spin accumulation at the F/N interface.

The device used to study the spin Peltier effect consists of a stack of two ferromagnetic layers separated by a layer of N (Fig. 2). Assuming that the spin relaxation length $\lambda_N$ is much larger than the thickness of N, we can neglect spin relaxation in N. A charge current $J_C$ is sent through the stack and the temperature at the top is anchored at $T_0$. When both ferromagnets are aligned parallel (P), the spin current is constant over the whole stack and there is no non-equilibrium spin-accumulation, i.e. $\mu_S = 0$ everywhere. The temperature follows a zigzag pattern[15], caused by the conventional (charge) Peltier effect (Fig. 2a). In the anti-parallel (AP) alignment, the situation is different. Now the spin current in the bulk of $F_1$ is opposite to the spin current in the bulk of $F_2$, leading to a spin accumulation at the interfaces[16]. According to Eq. 1, this gives rise to an additional temperature difference ΔT (Fig. 2b) for each F/N interface in the stack. Hence, the induced difference in temperature at the bottom layer between both magnetic configurations is now 2ΔT (Fig. 2c).

The specifically designed device used to study the spin Peltier effect is depicted in Fig. 3. It consists of a permalloy ($Ni_{80}Fe_{20}$)(Py) / copper / Py spin valve stack (150 x 80 $nm^2$ cross section) with a platinum (Pt) bottom contact and a gold (Au) top contact. Cross linked polymethyl methacrylate (PMMA) forms the insulating layer between these two contacts forcing the applied current through the spin valve stack. To probe the temperature of the device a constantan ($Ni_{45}Cu_{55}$) – Pt thermocouple is used, where constantan is chosen because of its large Seebeck coefficient (−32 $\mu$V/K, see Supplementary Table 1 in Supplementary E) . The thermocouple is electrically isolated from the bottom contact by an 8 nm thick aluminum-oxide layer, thereby excluding any spurious voltage pickup. In the measurement a current is sent from contact 1 to 2 through the stack, while recording the thermocouple voltage between contacts 3 and 4. For the 4-point spin valve signal measurements contacts 5 and 6 are used to probe the voltage. Using an ac lock-in measurement technique it is possible to separate the Peltier contribution (ΔT ∝ I) and the Joule heating contribution (ΔT ∝ $I^2$) by taking the first harmonic $V_1$ and second harmonic $V_2$ response, respectively[7,17,18]. The measurements are performed at room temperature.

The thermovoltage is recorded while sweeping the magnetic field in the in plane direction from negative to positive and back. Figure 4a shows the first harmonic response data with $R_1$ defined as $V_1$/I. The data shows four abrupt changes in $R_1$ when the magnetization of the Py layers switches from P to AP and back. A spin signal, ($R_P$–$R_{AP}$), of −80 μΩ is observed on top of a background signal, ($R_P$+$R_{AP}$)/2, of −0.44 mΩ. This corresponds to a temperature difference at the thermocouple between P and AP alignment of around 3 mK at 1 mA. Using the modeling described below we calculated that $2\Delta T = 7.6$ mK (Eq. 1). The second harmonic data is presented in Fig. 4b where $R_2=V_2/I^2$ and gives a spin signal of 110 mV A$^{-2}$ on a background of $-11.73$ V A$^{-2}$. This signal originates from the Joule heating[19] in the device and its change between the P and AP configuration (see Supplementary B).

The 4-probe spin valve signal shown in Fig. 4c gives a 100 mΩ spin signal. By matching the spin valve signal from our 3-D finite element model[13] to this measured value, we obtain a conductivity polarization ($P_\sigma$) of 0.61, which is close to the bulk value for Py[20,21]. The same model can now be used to extract the spin Peltier coefficient, $\Pi_S=\Pi_\uparrow-\Pi_\downarrow$, from the first harmonic measurement in Fig. 4a. The previously obtained value for $P_\sigma$ together with the electrical conductivities, thermal conductivities, spin relaxation lengths, the Peltier coefficients and Seebeck coefficients for each material are taken as input parameters (see Supplementary Table 1 in Supplementary E).

From the spin Peltier signal in Fig. 4a, we then obtain a spin Peltier coefficient for Py of $\Pi_S = -0.9$ mV. Similar results were obtained in two other devices giving values for the spin Peltier coefficient of $-1.1$ and $-1.3$ mV (see Supplementary C). The Thomson-Onsager relation $\Pi=S*T$ gives us a way to compare these values of $\Pi_S$ to the previously reported[7] spin-dependent Seebeck coefficient ($S_S$) of $-3.8$ µV K$^{-1}$. Using $\Pi_S = S_S*T$ and by taking T = 300 K, we get values for $S_S$ between $-3.0$ µV K$^{-1}$ and $-4.3$ µV K$^{-1}$ from our measurements, which is in agreement with ref. 7.

The observed background signal in the first harmonic measurement, due to the conventional Peltier effect, is a factor of 3 lower than we obtain from the modeling. We attribute this difference to difficulties with accurately determining the combination of Peltier effects for all the interfaces in the current path. Given these uncertainties we allow for the possibility that the actual value of the spin Peltier coefficient can be slightly higher than obtained from the current modeling.

To confirm that the spin Peltier effect is indeed the origin of the spin signal, the same device was measured at 77K (see Supplementary D). No spin signal could be observed in the $R_1$ measurement as expected from the temperature dependence of $\Pi_S \propto T^2$. This dependence is obtained by taking $S(T) \propto T$ in the Thomson-Onsager relation and gives an upper bound of $-0.1$ mV for $\Pi_S$ from the observed noise level at 77K.

In conclusion, we experimentally demonstrated a magnetically controllable heat current, driven by the spin-dependency of the Peltier coefficient. The relatively low efficiency of this effect in ferromagnetic metals restricts the cooling or heating power of the device. For use in applications the effect can possibly be enhanced by the use of nonmetallic materials. Promising candidates can be found in the ferromagnetic oxides[22,23] as high Seebeck coefficient have been reported[24]. These thermopower values are an order of magnitude larger than that in Py bringing the achievable temperature differences in the range of a few Kelvin. With electronic components becoming smaller and smaller the need for local and programmable refrigeration devices is growing and possibly the spin Peltier effect can fulfill this role.

Acknowledgement:

We would like to acknowledge B. Wolfs, M. de Roosz and J.G. Holstein for technical assistance and N. Vlietstra for initial pillar testing. This work is part of the research program of the Foundation for Fundamental Research on Matter (FOM) and supported by NanoLab, EU FP7 ICT Grant No. 251759 MACALO and the Zernike Institute for Advanced Materials.

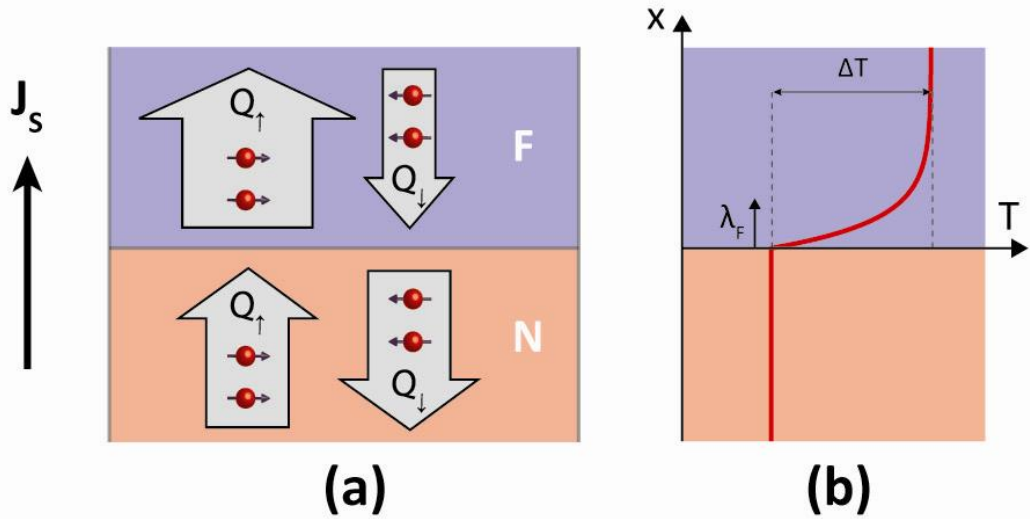

**Figure 1 Concept of the spin Peltier effect. a,** A pure spin current is sent through a nonmagnetic metal (N) / ferromagnetic metal (F) interface. In the N, the Peltier heat current for both spin species is equal. As the flow direction in the two spin channels is opposite, the total heat current is canceled. In ferromagnets, the heat currents are different for majority and minority carriers, leading to a net heat current from the interface into the ferromagnetic region. **b,** Generated temperature profile in the system. Spin relaxation in the ferromagnet reduces the spin current, thereby decreasing the induced heat current.

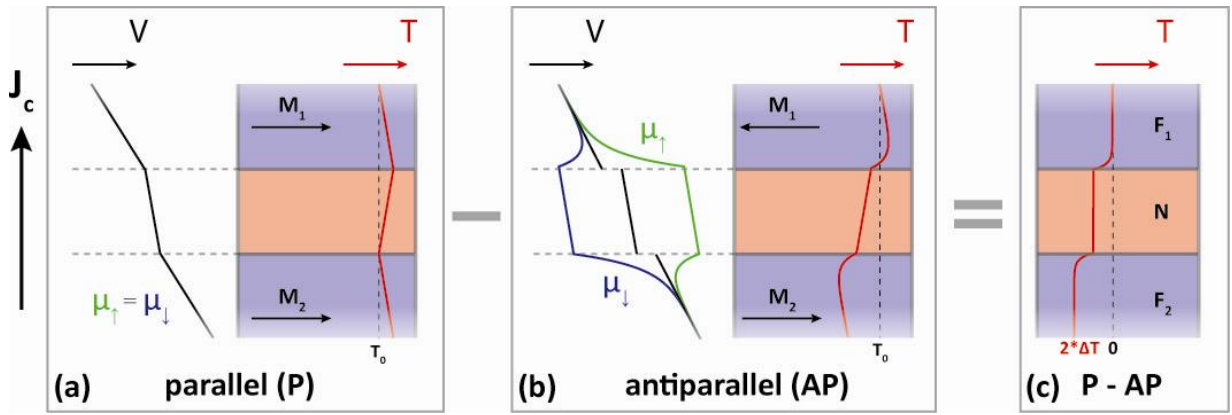

**Figure 2 Spin Peltier effect in a F/N/F spin valve stack.** Schematic figure showing the spin electrochemical potential and temperature throughout the stack for the parallel (P) and anti-parallel (AP) configuration of the ferromagnets. Spin-up/spin-down is defined as the spin direction of the majority/minority electrons in $F_2$. **a,** In the P situation, the spin current in the bulk of both ferromagnets equals the spin current in the N and no non-equilibrium spin-accumulation exists. Hence, there is no spin Peltier effect. **b,** For the AP configuration, the bulk spin currents in $F_1$ and $F_2$ are opposite. The resulting spin accumulation at the F/N interface leads to a spin Peltier contribution and hence, an altered temperature gradient. **c,** The spin Peltier effect causes the temperature at the bottom contact to change between the P and AP alignment.

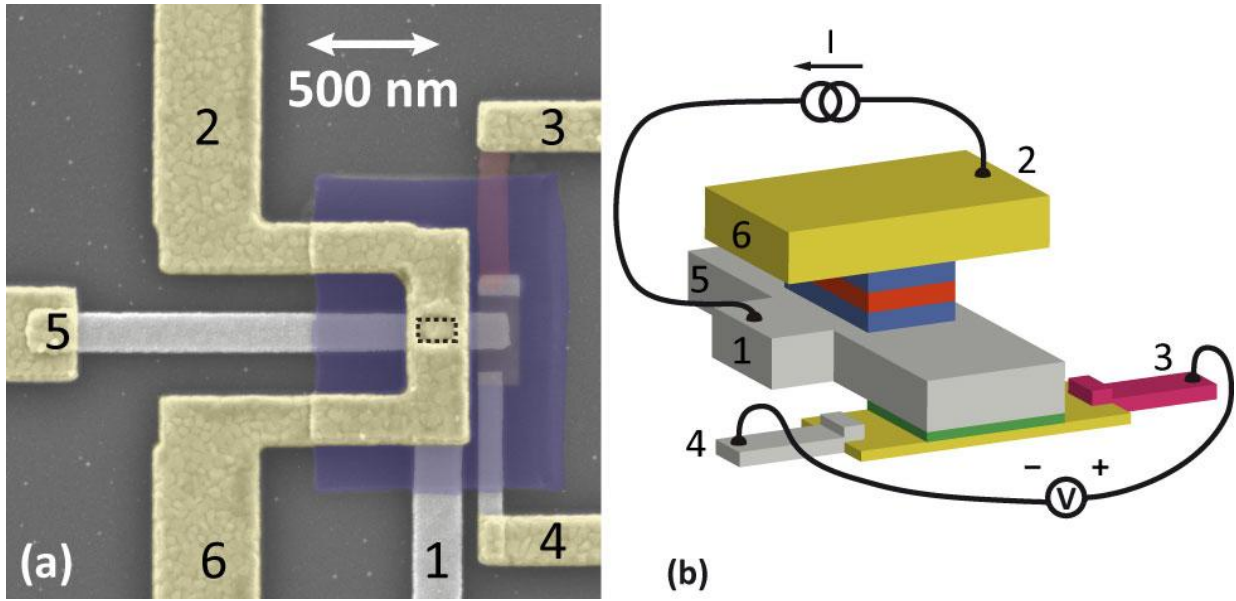

**Figure 3 Device geometry. a,** Scanning electron microscopy image of the measured device. The colors represent the different materials used. Yellow: Au top contact, grey: Pt bottom contacts, blue: cross linked PMMA, red: constantan ($Ni_{45}Cu_{55}$). **b,** Schematic representation of the device. Current is sent from contact 1 to 2, while recording the voltage between contacts 3 and 4. Contacts 1, 2, 5 and 6 are used for four probe spin valve measurements. The thermocouple is electrically isolated from the bottom contact by an $Al_2O_3$ (green) layer.

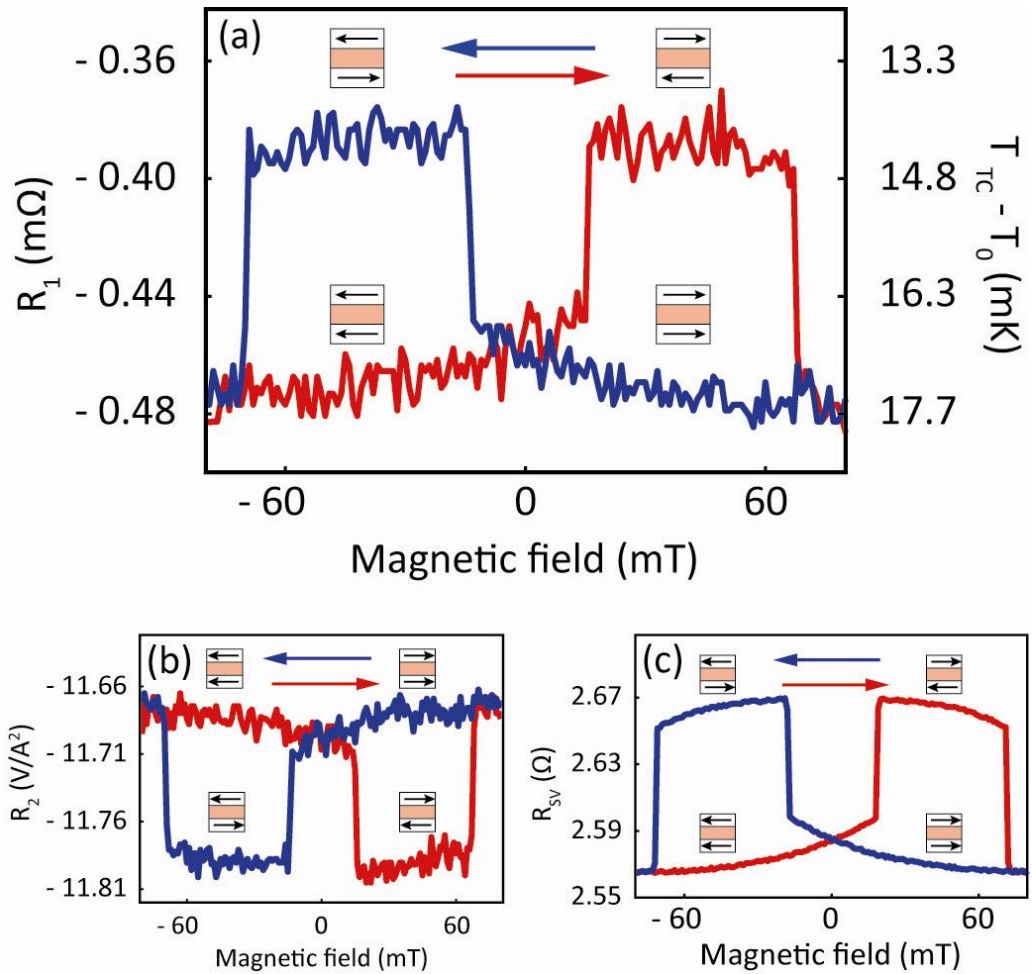

**Figure 4 Spin Peltier and spin valve measurements. a,** First-harmonic response signal, $R_1 \equiv V_1/I$, measured at the thermocouple with a root mean square current of 1 mA. Switches in $R_1$ are observed when the magnetization of the ferromagnetic layers changes from parallel to anti-parallel and back. On the right y-axis the temperature detected by the NiCu – Pt thermocouple relative to the reference temperature $T_0$ is given, using $T-T_0 = V_1/(S_{NiCu} - S_{Pt})$. **b,** Second-harmonic response signal, $R_2 \equiv V_2/I^2$, measured at the thermocouple. **c,** Spin valve measurement on the same device. $R_{SV}$ is determined by recording the 4-probe resistance of the stack using contacts 1,2, 5 and 6.

# Supplementary Information

## A. Calculation of the temperature gradient

We first derive an expression for the spin-dependent Peltier coefficient $\Pi_{\uparrow,\downarrow}$ in terms of the conventional Peltier coefficient $\Pi$ and the conductivity polarization $P_\sigma = \dfrac{\sigma_\uparrow - \sigma_\downarrow}{\sigma_\uparrow + \sigma_\downarrow}$. In the bulk of the ferromagnet $\nabla\mu_\uparrow = \nabla\mu_\downarrow = \nabla\mu_C$ and the Peltier heat current can be written as the sum of that of the separate spin channels, $-\Pi_\uparrow\sigma_\uparrow\nabla\mu_\uparrow - \Pi_\downarrow\sigma_\downarrow\nabla\mu_\downarrow = -\Pi\sigma\nabla\mu_C$, where we use $J_C = -\sigma\nabla\mu_C$ and $J_{\uparrow,\downarrow} = -\sigma_{\uparrow,\downarrow}\nabla\mu_{\uparrow,\downarrow}$ as the definitions of the electrochemical potentials $\mu_C$ and $\mu_{\uparrow,\downarrow}$. Using the spin-dependent conductivities $\sigma_{\uparrow,\downarrow} = \dfrac{\sigma}{2}(1 \pm P_\sigma)$, we obtain $\Pi = \dfrac{\sigma_\uparrow\Pi_\uparrow + \sigma_\downarrow\Pi_\downarrow}{\sigma}$. Rewriting the result gives us the relation for the spin-dependent Peltier coefficients:

$$\Pi_{\uparrow,\downarrow} = \Pi - \frac{1}{2}(P_\sigma \mp 1)\Pi_S \qquad (1)$$

where we define $\Pi_S = \Pi_\uparrow - \Pi_\downarrow$ as the spin Peltier coefficient.

Next, we derive an expression for the temperature gradient that develops in the ferromagnetic region for the general case when a spin current is accompanied by a charge current $J_C = J_\uparrow + J_\downarrow$. The Peltier heat current is given by $Q_\Pi = \Pi_\uparrow J_\uparrow + \Pi_\downarrow J_\downarrow$ and the temperature gradient is calculated by considering the total heat current in the ferromagnet $Q = Q_\Pi - \kappa\nabla T$, where $\kappa$ describes the thermal conductivity of the electron and phonon system. For simplicity, we assume that no heat can enter or leave the stack ($Q = 0$) and disregard Joule heating. Then we can write $\nabla T = -\frac{1}{\kappa}[\Pi_\uparrow \sigma_\uparrow \nabla\mu_\uparrow + \Pi_\downarrow \sigma_\downarrow \nabla\mu_\downarrow]$ and from the definition of the spin-dependent conductivity and Peltier coefficient, we find:

$$\nabla T = -\frac{\sigma}{\kappa}\left(\Pi \nabla\mu_C + \frac{1}{4}(1-P_\sigma^2)\Pi_S \nabla\mu_S\right) \qquad (2)$$

with $\Pi$ is the spin-independent Peltier coefficient and $\mu_S = \mu_\uparrow - \mu_\downarrow$ the spin accumulation. The electrochemical potential is here derived from current conservation $J_C = J_\uparrow + J_\downarrow = -\sigma\nabla\mu_C$ and by substitution of $J_{\uparrow,\downarrow}$ with $-\sigma_{\uparrow,\downarrow}\nabla\mu_{\uparrow,\downarrow}$, we write $\nabla\mu_C = \frac{\sigma_\uparrow \nabla\mu_\uparrow + \sigma_\downarrow \nabla\mu_\downarrow}{\sigma}$.

The first term of Eq. 2 describes the conventional Peltier effect in the absence of spin accumulation. The second term describes what happens if a spin accumulation is present in the ferromagnet. According to Eq. 2, this gives rise to an additional temperature gradient which depends exclusively on the gradient of the spin accumulation in the ferromagnetic layer and is therefore magnetically controllable. Since spin relaxation forces the spin accumulation to decrease exponentially in the ferromagnetic region, we can write $\mu_S = \mu_S^0 \exp(-x/\lambda_F)$ with $\lambda_F$ the spin relaxation length. The conventional Peltier term leads to a constant temperature gradient independent of the spin accumulation. By integrating only the spin Peltier term of Eq. 2, we obtain the temperature difference between the F/N interface and the bulk of the ferromagnet:

$$\Delta T = \frac{\sigma}{4\kappa}(1-P_\sigma^2)\,\Pi_S \mu_S^0 \qquad (3)$$

where $\mu_S^0$ is the spin accumulation at the interface. Here we find that the induced temperature drop depends directly on the spin accumulation at the F/N interface.

## B. Second harmonic response (Joule heating)

The second harmonic response signal (see Supplementary Fig. 1) originates from Joule heating in the device and is proportional to $I^2 R$. This dependence on R causes a change in Joule heating[13,17,19] when the resistance of the spin valve stack changes from the P to AP configuration and vice versa. Changes in Joule heating in the spin valve stack are picked up by the thermocouple and show up in the second harmonic response measurement, $R_2$, as they depend on $I^2$. In our model we explicitly take in to account the heat generation due to energy dissipation related to spin relaxation[19]. From the model we then obtain the background and spin signal, which are approximately two times higher than observed in the measurement. We explain this by inefficiency in the temperature sensing, owing to the discrepancy between modeling parameters and the actual, experimental values. Moreover, a big part of the background Joule heating takes place in the Pt bottom contact. The cross linked PMMA, not included in the modeling, covers this contact thereby lowering the background Joule heating signal.

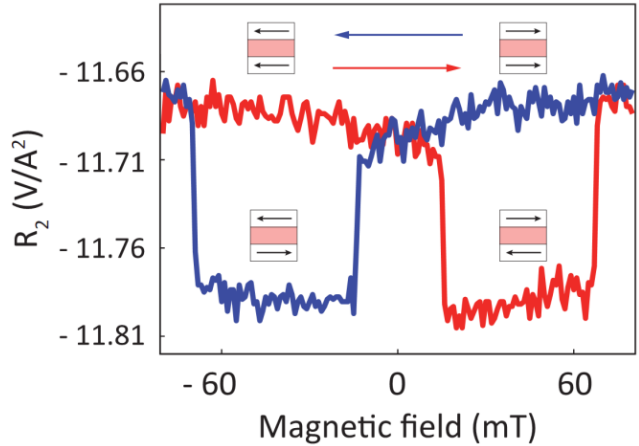

**Supplementary Figure 1 Joule heating measurement.** Second-harmonic response signal, $R_2 \equiv V_2/I^2$, measured at the thermocouple with a root mean square current of 1mA.

## C. Results for two other samples

The spin Peltier measurements were performed on two other samples of the same batch and are presented in Supplementary Fig. 2. The first sample (Supplementary Fig. 2a) shows a spin Peltier signal of −100 µΩ on a background of −0.55 mΩ and the second (Supplementary Fig. 2b) a −110 µΩ spin Peltier signal on a −0.56 mΩ background. These values are somewhat higher than for the sample discussed in the main text. The observed variation can be attributed to a slightly higher efficiency of the thermocouple of these samples and/or small differences in thermal anchoring, aluminum oxide thickness and lithographic alignment. The switching that is observed prior to sweeping through zero field is due to interaction between the magnetic dipole fields of the two Py layers, which favors an AP alignment. The sample to sample variation of the switching field position has been seen in several batches for different experiments and can be attributed to for instance small variations in cross section of the pillar. Extracting the spin Peltier coefficient from this data in the same way as discussed in the main text gives values for $\Pi_S$ of −1.1 and −1.3 mV.

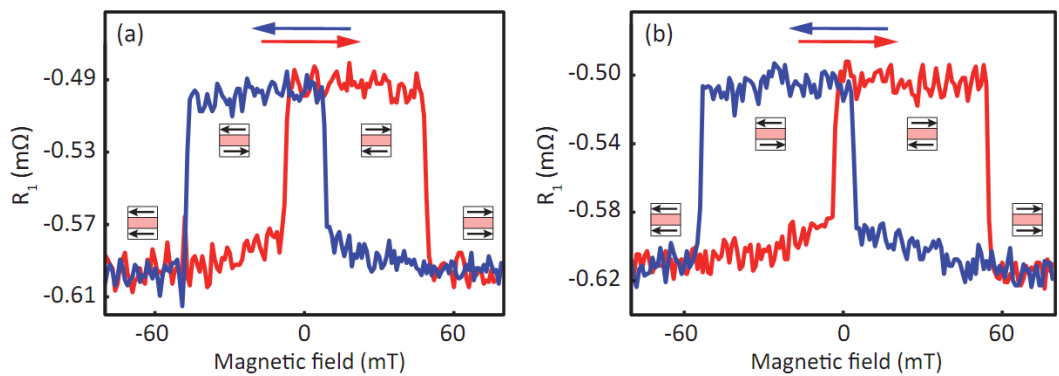

**Supplementary Figure 2 Spin Peltier measurements for two other samples.** First-harmonic response signal, $R_1 \equiv V_1/I$, measured at the thermocouple with a root mean square current of 1 mA. In **a** the results for sample 2 are shown and in **b** those for sample 3.

## D. <u>Measurements at 77K</u>

The presented measurements were repeated on the same sample at liquid nitrogen temperature (77K). This was done to confirm that the first harmonic spin signal is indeed caused by the spin dependency of the Py Peltier coefficient. From the Thomson-Onsager relation, $\Pi_{\uparrow,\downarrow} = T*S_{\uparrow,\downarrow}$, together with the fact that the Seebeck coefficient shows a dependency on temperature, it becomes clear that $\Pi_S$ (Supplementary A) and thereby the spin Peltier effect will decrease when lowering the temperature. The spin Peltier measurement at 77K is presented in Supplementary Fig. 3a and shows no difference between P and AP alignment. The disappearance of the spin signal at low temperature supports our conclusion that the room temperature spin signal can be attributed to the spin Peltier effect. At the same time the background signal, which originates from the conventional Peltier effect, remains almost the same. This can be explained by the fact that for the spin Peltier effect only the Peltier coefficient of Py plays a role whereas for the Peltier background the difference between all the Peltier coefficients in the current path are important. The Peltier coefficient is proportional to the Seebeck coefficient ($\Pi=S*T$) whose temperature dependence does not have to be the same for different materials. Together with a change in thermal conductance between different temperatures it is possible for the regular Peltier effect contribution to not show a decrease when going from room temperature to 77K.

The spin valve measurement shown in Supplementary Fig. 3c shows a decrease in background resistance due to an increase of the conductivities at lower temperatures. The bigger spin signal that is observed is caused by the spin relaxation lengths increasing with lowering of the temperature.

As the Joule heating depends on the resistance, the increase of the materials' conductivities at 77K will give a lower second harmonic background signal, which is in accordance with the measurement shown in Supplementary Fig. 3b. At the same time the second harmonic spin signal goes up because of the increased difference in resistance between P and AP alignment shown in the spin valve measurements. In the measurement this increase is smaller due to temperature dependences of the Seebeck coefficients and thermal conductivities.

In conclusion we can say that the disappearance of the first harmonic signal, while the spin valve signal increases, rules out the possibility of it originating from spin valve voltage pick up and is consistent with the spin Peltier effect. Furthermore the second harmonic and spin valve measurement behavior confirm the lowering of the reference temperature and the correct operation of the device and thermocouple.

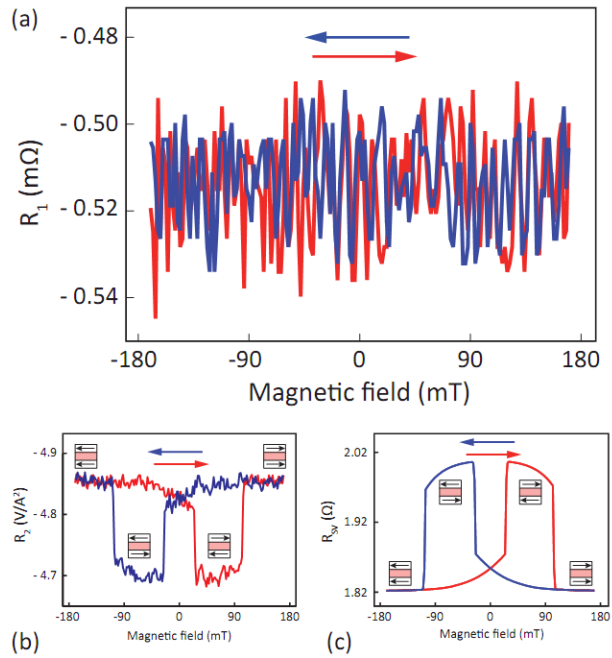

**Supplementary Figure 3 Measurements at 77K. a,** First-harmonic response signal, $R_1 \equiv V_1/I$, at 77K measured at the thermocouple with a root mean square current of 1 mA. **b,** Second-harmonic response signal, $R_2 \equiv V_2/I^2$, at 77K measured at the thermocouple with a root mean square current of 1 mA. **c,** Spin valve measurement at 77K on the same device.

E. **Modeling parameters**

$T_0 = 300K$

| Material | σ [S m$^{-1}$] | Π [mV] | λ [nm] | κ [W m$^{-1}$ K$^{-1}$] |
|---|---|---|---|---|
| **Au** | $2.2 \times 10^7$ | 0.51 | 80 | 300 |
| **Pt** | $9.5 \times 10^6$ | -1.5* | 5 | 72 |
| **Cu** | $4.3 \times 10^7$ | 0.48 | 350 | 300 |
| **Py** | $4.3 \times 10^6$ | -6.0* | 5 | 30 |
| **NiCu** | $2.0 \times 10^6$ | -9.6* | 5 | 20 |
| **SiO$_2$** | $1.0 \times 10^{-13}$ | 0 | - | 1 |
| **Al$_2$O$_3$** | $1.0 \times 10^{-13}$ | 0 | - | 30 |

**Supplementary Table 1 Input parameters for the modeling**.

\* The Peltier coefficient was determined in a separate device specifically designed to accurately determine the Seebeck/Peltier coefficient of a material.